\documentstyle[12pt]{article}
\newtheorem{theorem}{Theorem}

\newcommand{\CMP}[3]{Commun. Math. Phys. {\bf #1} (#2) p. #3 }

\title{Interacting Quantum Fields on a Curved Background\thanks{to
    appear in the Proceedings of the ICMP Brisbane 1997}} 
\author{Romeo Brunetti\\ Dipartimento di Scienze Fisiche\\ Universit\'a 
di Napoli\\ I-80125 Napoli, Italy \and 
Klaus Fredenhagen \\ Universit\"at Leipzig\\ and \\ MPI for 
Mathematics in the Sciences\\ D-04103 Leipzig, Germany
\thanks{on leave of absence from II. Institut f\"ur 
Theoretische Physik, Universit\"at Hamburg, D-22761 Hamburg, Germany}} 
\date{}
\begin{document} 
\maketitle 
\begin{abstract} 
 A renormalized perturbative expansion of
 interacting quantum fields on a globally hyperbolic spacetime is performed by
 adapting the Bogoliubov Epstein Glaser method to a curved background.  The
 results heavily rely on techniques from microlocal analysis, in particular on
 Radzikowski's characterization of Hadamard states by wave front sets of 
 Wightman functions. 
\end{abstract} 

Quantum field theory on a
curved background is supposed to describe correctly the influence of a
gravitational field on quantum fields as long as the relevant length scales are
much larger than the Planck length. The
modifications compared to quantum field
theory on Minkowski space are mainly due to some nonlocal features in the
standard formulation of quantum field theory. The
most important ones are the spectrum
condition and the existence of a vacuum, properties expressing stability of the
system and, on a more technical side, making possible the transition to a
euclidean formulation.  Even for free fields the changes are by no means
trivial, leading to new phenomena as e.g. the
Hawking radiation of black holes.
It seems now to be clear that the notion of a vacuum has to be replaced 
by the choice of a class of admissable states
\cite{HNS} which was identified to be the class of Hadamard
states \cite{dWB,Kay-Wald,Verch}.  Accordingly, also the Feynman propagator
is no longer unique, but its singular part is
fixed  and has essentially the same 
form as on Minkowski space.  Therefore the ultraviolet
divergences of perturbation theory should be treatable and should be similar to
those in Minkowski space. This expectation fits
with the results on renormalization
of euclidean theories on Riemannian spaces \cite{Lue,Bunch}. It is not obvious,
however, how these results can be used for renormalization on lorentzian
manifolds.

For a rigorous treatment one needs an effective mathematical framework.  It was
first observed by Radzikowski \cite{R} that microlocal analysis, in particular
the notions of wave front sets and Fourier Integral Operators \cite{H,DH} are
ideally suited to deal with the singularities of propagators on curved
spacetimes.

We choose the Bogoliubov-Epstein-Glaser method
\cite{EG}  for the construction of
interacting fields.  This method can be formulated locally and amounts to the
extension of time ordered n-point distributions to
coinciding points,  a problem which can 
nicely be treated by the techniques of microlocal analysis.  But an important
technical ingredient in the Epstein-Glaser method is translation invariance,
therefore one of the main problem which had to be
solved was a  generalization of
this method to a situation where invariance is
replaced by a suitable  smoothness
condition.  A similar problem occurs in the presence of external fields
\cite{DM}.

Let us start from a free neutral scalar field $\varphi$ which satisfies the
Klein Gordon equation on a globally hyperbolic spacetime ${\cal M}$ with a 
metric tensor $g$.  The commutation relations are 
  \[ [\varphi(x),\varphi(y)]=i\Delta(x,y) \] 
where $\Delta$ is the difference between
the (unique) retarded and advanced Green function of the Klein Gordon operator.
According to Radzikowski \cite{R} a Hadamard state $\omega$ on the algebra 
generated by $\varphi$ is a quasifree state whose 2-point function is a 
bisolution 
of the Klein-Gordon equation with antisymmetric part $i\Delta$ and whose wave 
front set is the positive frequency part of the
wave front set of $\Delta$.  It was 
shown in \cite {BFK} that as an immediate consequence Wick polynomials can be 
defined as operator valued distributions on a
dense  invariant subspace of the GNS
Hilbert space of $\omega$. 

The time ordered products of Wick polynomials are well defined as symmetric 
operator valued distributions on testfunctions $f$
with  support on noncoinciding points,
   \[ \mbox{supp}f\subset \{(x_1,\dots,x_n),
   x_i\in{\cal M},  x_i\ne x_j,i\ne j\}. \]
They have an expansion into sums of pointwise
products of  numerical distributions and 
multiple Wick products. The numerical
distributions are  time ordered functions, i.e.
expectation values of time ordered 
products of sub-Wick polynomials. The problem
amounts  now to the extension of these 
time ordered functions to coinciding points such
that the pointwise  products remain
well defined.

The first step is to guess the wave front set of
these extensions.  We found a condition 
on these wave front sets which allows to build the
mentioned  pointwise products.
One then makes an inductive construction of time
ordered functions.  Finally, one defines 
interacting fields as formal power series of operator valued distributions.

The wave front set of a distribution on a manifold
is the subset  of the cotangent 
bundle which characterizes for every point of the
manifold  the directions of nonrapid 
decrease of a local Fourier transform,
\begin{eqnarray*}
        &\mbox{WF}(f)=\{(x,k)\in \mbox{T}^*{\cal M},k\neq 0,   &\\
        &\hat{\psi f} \mbox{ not fast decreasing near } k
        \ \forall \psi \in {\cal D}({\cal M}) \mbox{ with } \psi(x) \neq 0\}.&
\end{eqnarray*}
The Fourier transform is defined within a suitable
chart at $x$, but the wave front set is
independent of the choice of the chart. 

For the Feynman propagator 
\[ \Delta_F^{\omega}(x,y)=\omega(T(\varphi(x)\varphi(y))\]
associated to a Hadamard state $\omega$ it is
\[ \mbox{WF}\Delta_F^{\omega}=\{(x,x';k,k')\in
\mbox{T}^*{\cal M}^2 \setminus \{0\},
(x,k) \sim (x',-k')\}, \]
where $(x,k)\sim (x',k')$ means that either $x=x'$
and $k=k'$ or  there is a null 
geodesic from $x$ to $x'$, $k$ is coparallel to
the  tangent vector of the geodesic at 
$x$, $k'$ is the parallel transport of $k$ along
the geodesic  and $k\in V_{\pm}$ if the 
geodesic is future (past) directed.

The first problem is to find a suitable condition
on the wave front set of  time ordered
functions. We are looking for an expansion of the form
\[ T:\!\varphi(x_1)^{\alpha_1}\!:\cdots :\!\varphi(x_n)^{\alpha_n}\!:
=\sum_{\beta \leq \alpha}
t_{\beta}^{\alpha}(x_1,\ldots,x_n) :\!\ varphi(x_1)^{\beta_1}
\cdots \varphi(x_n)^{\beta_n} \!: \]
with
\[ t_{\beta}^{\alpha}(x_1,\ldots,x_n)={\alpha \choose \beta} 
\omega(T:\!\varphi(x_1)^{\alpha_1-\beta_1}\!:\cdots
:\!\varphi(x_n)^{\alpha_n-\beta_n}\!:),
\]
$\alpha,\beta$ multiindices, where the time
ordered products are  symmetric under 
permutations of $\{1,\ldots,n\}$ and coincide with
the operator product  if the points
$x_1,\ldots,x_n$ can be time ordered,
\[x_i \notin {\cal J}_-(x_{i+1}), i=1,\ldots ,n-1 , \]
${\cal J}_-(x)$ denoting the causal past of $x$.
Since the Wick products are local and relatively
local,  this condition is compatible
with the symmetry requirement.
The two conditions mentioned above fix the time ordered products on a globally
hyperbolic manifold uniquely on the set of
noncoinciding points.  The wave front set of 
the time ordered functions can be determined from
the wave  front set of the Feynman
propagator and turns out to be contained in the set
\[
\begin{array}{ccc}
\Gamma_n^{\mbox{to}}&=&\{(x_1,\ldots,x_n;k_1,\ldots,k_n)\in 
\mbox{T}^*{\cal M}^n \setminus \{0\};\mbox{there is a graph } G\\
& &\mbox{with vertices } \{1,\ldots,n\}, \mbox{an association of lines } l 
\mbox{ of the graph to}\\
& &\mbox{ future directed lightlike geodesics } \gamma_l \mbox{ from }
 x_{s(l)} \mbox { to } x_{r(l)}\\
& &\mbox{ with covariantly constant coparallel covector fields } k_l \mbox{ 
on } \gamma_l \\
& &\mbox{with values in the closed forward lightcone such that }\\
& & k_i=\sum_{s(l)=i} k_l(x_i) -\sum_{r(l)=i} k_l(x_i) \quad \}
\end{array}
\]
We now start an inductive construction of the time
ordered products  by postulating that
also the wave front set of the extended time ordered functions is contained in 
$\Gamma_n^{\mbox{to}}$. Note that since coinciding
points can  trivially be connected by 
future oriented lightlike geodesics, one gets a
restriction  only on the sum of covectors
at coinciding points. This may be considered as a
local  version of translation invariance.

The above condition on the wave front set  leads to the following version of 
Epstein-Glaser's Theorem 0:
\begin{theorem}
Let $t \in {\cal D}'({\cal M}^n)$ with
$\mbox{WF}(t)  \subset  \Gamma_n^{\mbox{to}}$. Then
the pointwise products
\[
t(x_1,\ldots,x_n):\!\varphi(x_1)^{\beta_1}\cdots
\varphi(x_n)^{\beta_n}  \!: \]
are well defined operator valued distributions
with an invariant  domain of definition.
\end{theorem}
 
We now assume that all time ordered products of  $n'<n$ factors have been 
constructed and satisfy the requirements of
symmetry and  unitarity, admit the expansion
above with $\mbox{WF}(t)\subset
\Gamma_{n'}^{\mbox{to}}$ and  fulfil the factorization 
property 
\[ T \prod_{i=1}^{n'} :\! \varphi(x_i)^{k_i}\!:= 
   T \prod_{i=1}^l :\! \varphi(x_i)^{k_i}\!:
   T \prod_{i=l+1}^{n'} :\! \varphi(x_i)^{k_i}\!: \]
if $x_i \notin {\cal J}_-(x_j)$ for $i=1,\ldots,l,\quad j=l+1,\ldots,n'$.

Then on ${\cal M}^n\setminus {\rm D}_n$, 
${\rm D_n}=\{(x,\ldots,x)\in {\cal M}^n,x\in {\cal M}\}$  
denoting the total diagonal, the time ordered products are uniquely defined 
and satisfy the previous conditions.

It remains to extend the time ordered functions to
the diagonal.  On Minkowski space
one uses translation invariance to eliminate one 
coordinate so that one needs to extend the
distribution  only to a single point. 
The singularity at this
point  (taken to be the origin) can conveniently
be  desribed in terms of Steinmann's
scaling degree \cite{ST}
\[\mbox{sd}(t)=\mbox{inf}\{\delta,
\lambda^{\delta}t(\lambda \cdot) \rightarrow 0,
\lambda \rightarrow 0 \},\]
and one obtains the possible extensions with the same scaling degree 
by decomposing the space of test functions into a direct sum of the 
space of functions which vanish at the origin with order 
$\mbox{sd}(t)-4(n-1)$ and a complementary (finite dimensional) subspace: on 
the first summand there is a unique extension with the same scaling 
degree whereas on the second summand one may choose an arbitrary 
linear functional. The singularity degree $\mbox{sd}(t)-4(n-1)$ coincides 
with the degree of divergence obtained by the usual power counting 
rules.

On a curved spacetime one may introduce near the diagonal center of 
mass and relative coordinates in terms of the exponential function,
\[(x_{1},\dots,x_{n})=(\exp_{x}\xi_{1},\dots,\exp_{x}\xi_{n}),
\quad \sum\xi_{i}=0,
\xi_{i}\in\mbox{T}_{x}{\cal M},\]
and perform the extension only with respect to the relative coordinates. 
But the nontrivial dependence on the center of mass coordinate 
requires a careful treatment for which techniques from microlocal 
analysis turn out to be useful; one can introduce the concept of a 
scaling degree relative to a submanifold (here the diagonal) whose 
tangent bundle is orthogonal to the wave front set, and one finds that 
the inductive computation of this scaling degree gives the same 
results as for the scaling degree in the translationally invariant 
situation. For details see \cite{BF2}.

After the construction of time ordered products of Wick polynomials 
one can define interacting fields in the sense of formal power series 
by Bogoliubov's formula
\[\varphi_{g}^k(x)=\frac{\delta}{\delta 
h_{k}(x)}S(g)^{-1}S(g+h)_{|h=0},\]
where the ``S-matrix'' $S(g)$ is defined by
\[S(g)=\sum_{n=0}^{\infty}\frac{i^n}{n!}\sum_{k_{1},\dots,k_{n}}\int 
dx_{1}\cdots dx_{n}T\prod_{j=1}^n 
:\!\varphi(x_{j})^{k_{j}}\!:g_{k_{j}}(x_{j})\]
with test functions $g_{j}$. 
These fields are local and satisfy field equations. They 
depend only on the values of $g$ in the
past. Moreover,  within a fixed causally 
closed region ${\cal O}$  their dependence on the values of $g$ 
outside of ${\cal O}$ is described by a unitary transformation. Hence 
the structure of the local algebras of observables generated by the 
interacting fields is independent of the values of the couplings $g$ 
outside of the region one is interested in; one therefore obtains a 
purely local construction (in the sense of formal power series) of the 
Haag-Kastler net of the interacting theory. On a curved space time, 
this enables us to ignore space time singularities outside the region 
we are interested in, but also on Minkowski space this local 
construction may be useful for separating the infrared problems from 
the ultraviolet problems.

%%%%%%% 
\end{document}